# Bošković's method for determining the axis and rate of solar rotation by observing sunspots in 1777

**Mirko Husak[1] · Roman Brajša[2] · Dragan Špoljarić[3] · Davor Krajnović[4] · Domagoj Ruždjak[2] · Ivica Skokić[2] · Dragan Roša[5] · Damir Hržina[5]**

**Abstract** In 1777 Ruđer Bošković observed sunspots, determined their positions and the solar rotation elements by his own methods briefly described here. We repeat his calculations of the mean solar time, sunspot positions, and solar rotation elements using both the Bošković's original equations and equations adapted for modern computers. We repeat the calculations using two values of the obliquity of the ecliptic, Bošković's, and an interpolated one. Using his 1777 observations, Bošković obtained the solar equator's inclination, ecliptic longitude of the ascending node, and sidereal and synodic rotation periods. We analyzed and compared these original Bošković results with our repeated calculations. Our results confirm the validity of Bošković's methods and the precision of his calculations. Additionally, the paper presents the solar differential rotation determination using all 1777 observations by Ruđer Bošković.

✉ M. Husak
mhusak@geof.hr

[1] Trakošćanska 20, 42000 Varaždin

[2] Hvar Observatory, Faculty of Geodesy, University of Zagreb, Kačićeva 26, 10000 Zagreb, Croatia

[3] Faculty of Geodesy, University of Zagreb, Kačićeva 26, 10000 Zagreb, Croatia

[4] Leibniz-Institut für Astrophysik Potsdam (AIP), An der Sternwarte 16, 14482 Potsdam, Germany

[5] Zagreb Astronomical Observatory, Opatička 22, 10000 Zagreb, Croatia

# 1. Introduction

Ruđer Bošković[1] (Dubrovnik, 1711 — Milan, 1787) was a Croatian Jesuit priest. He was educated at the *Colegium Ragusinum* in Dubrovnik (Ragusa) and then *Colegium Romanum* in Rome, where he later became a professor of mathematics. He published many works in various fields of mathematics, physics, astronomy, geodesy, cartography, archeology, civil engineering, and philosophy.

He made numerous constructions and improvements of optical instruments, especially after his visit to London in 1760. Having accepted to take a chair of Mathematics at the University of Pavia in 1764, he was also responsible for establishing the Brera Observatory, located in present-day Milan, Italy. He substantially improved Brera astronomical observatory, where he implemented new instruments which he had seen during his stay in London in 1760. In 1761, he was elected as a member of The Royal Society in London (James, 2004; MacDonnell, 2014).

From 1770 he became a lecturer in Milan. After the abolition of the Jesuits in 1773, he went to France, where he obtained French citizenship and was appointed director of Naval Optics of the French Royal Navy. In 1782 he requested and received a leave in Bassano, Italy, where he planned to collect and publish his complete scientific works. In 1785, he published *Opera pertinentia ad opticam et astronomiam*, the main source for the present work (Boscovich, 1785a).

## 1.1. Bošković's contribution to astronomy

Bošković made important contribution to natural science, especially in the field of astronomy. He investigated theoretically and observationally, and he mathematically applied his own methods to many astronomical phenomena: solar rotation and sunspots, transit of the planet Mercury across the Sun, the aberration of stars, eclipses, cometary and planetary orbits, among others. His dominant interest in astronomy is presented in the five-book compendium of his works *Opera pertinentia ad opticam et astronomiam*, written mostly in Latin and partly in French (Boscovich, 1785a).

Bošković wrote the complete scientific experiment on determination of solar rotation elements. The chapter *Opuscule II* consists of descriptions of methods, observations, equations, numerical example, discussion of Bošković's results for the first sunspot and the results known from his time, as well as an appendix with all his observations of four sunspots in 14 days, September 12-29, 1777 (Boscovich, 1785b). Ruđer Bošković wrote the *Opuscule II* of the fifth tome in French in the honor of French king Louis XVI.

[1] In Croatia we write his name as Ruđer Bošković, but in his works we can find it written as follows:

Rogerii Josephi Boscovich in *Opera...* (Boscovich, 1785a)
Rogerius Joseph Boscovich in works *Philosophical Transcations* (Boscovich, 1777)
Rogerio Josepho Boscovich in *Philosophiae Naturalis...* (Boscovich, 1763)
Rogerii Josephi Boscovich in *De solis ac lunae defectibus ...* (Boskovic, 1760).

.

### 1.2. Works on solar rotation: Observing sunspots by Bošković in 1777

His first scientific work, so-called dissertation, *De maculis solaribus* in Latin (On sunspots) deals with his own methods for determination of solar rotation elements based on observed sunspot positions on the solar disk (Boscovich, 1736). His second work on sunspots, *Opuscule II*, is a complete experiment for determination of the solar rotation elements. It comprises a description of his methods, equations and accompanied drawings, a description of instruments used for his observations and measurements, a numerical example and a detailed calculation description, and finally a discussion of his results in the historical context of that time (Boscovich, 1785b).

This paper presents the main results of a research titled *Contribution of Ruđer Bošković in Determination of Solar Rotation Elements Using Sunspot Observation.* Interpretation and reconstruction of the original Bošković's methods and his calculation examples from *Opuscule II* was the goal of our previous research (Husak, Brajša, and Špoljarić, 2021a, Husak, Brajša, and Špoljarić, 2021b and Roša et al., 2021). In the present work, we validate his methods and results by repeating the calculations, using both his original and modernized equations.

### 1.3. Works on solar rotation by observing sunspots

Telescope observations of the Sun started in 1610 when Galileo observed sunspots and came to the conclusion that Sun rotates with approximately 30-day period. *The Beginning of Modern Solar science, 1610–1810* presents the telescope observation period from its beginning from a historical point of view (Hufbauer, 1993). We can find a description of the Sun, observing the solar surface, sunspot fine structures, physics from drawings (the Wilson effect, solar rotation, and sunspot areas), and solar diameter determination, in the monograph by Vaquero and Vázquez (2009) as well as telescopic observations before, during and after Maunder Minimum and during the Dalton Minimum.

Clette et al. (2023) describes the importance of recalibration of the Sunspot Number during the 11-year cyclic and secular variation of solar activity. This is especially important for the study of secular variation of solar activity. In this context, it is important to investigate all available historic observations of sunspots.

Many papers deal with sunspot positions on the apparent solar disk using historical sunspot observations by Thomas Harriot (Herr, 1978), Galileo Galilei, Cigoli (Ludovico Cardi), Christoph Scheiner and Monreali Sigismondi di Colonna (Vokhmyanin, Arlt, and Zolotova, 2021), Christoph Scheiner (Arlt et al., 2016) and Barnaba Oriani (Nogales et al., 2020). Historical sunspot records (observations, drawings, and measurements) from the pretelescopic era, the telescopic period that includes the Maunder and Dalton minima, and the period of photographic records up to 1900 were reviewed by many researchers. This led to a better understanding of sunspots over time and physical quantities derived from sunspot positions (Arlt and Vaquero, 2020). Less work is, however, devoted to understanding the solar differential rotation and rotation elements determined from historical records (Royal, 1924; Eddy, 1976; Eddy, Gilman, and Trotter, 1977; Yallop et al., 1982; Schröter, 1985; Casas, Vaquero, and Vazquez, 2006; Vaquero, 2007; Arlt and Fröhlich, 2012; Sudar and Brajša, 2022).

The present work deals with results of the research on determination of solar rotation elements using sunspot positions on apparent solar disk observed by Ruđer Bošković in Sens nearby Paris in 1777.

Analysis and discussion deals with Bošković's sources (de Lalande and Dupius, 1771) for his discussion *Réflexions* (Boscovich, 1785b) and a contemporary source with historical solar rotation determinations (Wöhl, 1978, Table 1). The source for the present work is Britannica (1841)[2] published earlier than the sources used today (Carrington, 1863; Spörer, 1874).

Many scientists used drawings of the solar disc in the historical sunspot records, but modern solar drawings have their place in contemporary solar research (Vaquero and Vázquez, 2009). Ruđer Bošković determined the axis and rate of solar rotation by observing sunspots, but in his research there are no drawings of the solar disk. This research includes all sunspot observations of Bošković as we know.

## 2. The methods for solar rotation elements determination by Ruđer Bošković

Ruđer Bošković described his own methods for determination of sunspot positions and for solar rotation elements: the ecliptic longitude of the ascending node $\Omega$, the inclination of the solar equator to the ecliptic $i$, and the rotation periods, sidereal $T'$, and synodic $T''$. Bošković uses the letter $N$ for the ecliptic longitude of the ascending node, today we use $\Omega$.

His methods are based on sunspot positions determined from measurements at the apparent solar disk. Using three positions of the same sunspot, he derived the ecliptic longitude of the ascending node, the solar equator's inclination to the ecliptic, and sidereal and synodic rotation periods. He repeated the calculations using different pairs and/or triplets to obtain the arithmetic means of the solar rotation elements. His methods and methodology are described as follows.

### 2.1. The method for $\Omega$

The intersections of the ecliptic and solar equator determine the ascending node ☊ and the descending node ☋ on certain ecliptic longitudes. The method determines the ecliptic longitude of the maximum ecliptic latitude of a sunspot using two positions of the same sunspot at equal ecliptic latitudes. Observing two positions of the same sunspot at equal ecliptic latitudes is practically almost impossible, but mathematically Bošković simulates two positions at the same ecliptic latitudes using three positions, two on the left-hand side, and one on the right-hand side of the maximal ecliptic latitude of the sunspot (Figure 1) (Boscovich, 1785b, §.IV., №45). The ecliptic longitude of the ascending node is the ratio of differences of three sunspot positions in ecliptic coordinates $S(B,C)$, $S'(B',C')$, and $S''(B'',C'')$. Two sunspots at the same ecliptic latitude determine maximal sunspot ecliptic latitude, which is the ecliptic longitude of the ascending node Ω increased by 90° (Equation 1):

$$\Omega = N = f_N(B, B', B'', C, C', C''),$$

$$\Omega = N = \frac{1}{2} \cdot [(B + B') + (B'' - B') \cdot (C - C')/(C'' - C')] \pm 90^\circ. \tag{1}$$

[2] *CHAP. II. OF THE SUN. The sections: Sect. I. – Of the Apparent Circular Motion of the Sun in the Ecliptic, and Position of the Ecliptic in Space.* (pp. 764–768) and *Sect. V. – Of the Spots of the Sun, his Rotation, and Constitution.* (pp. 778–784).

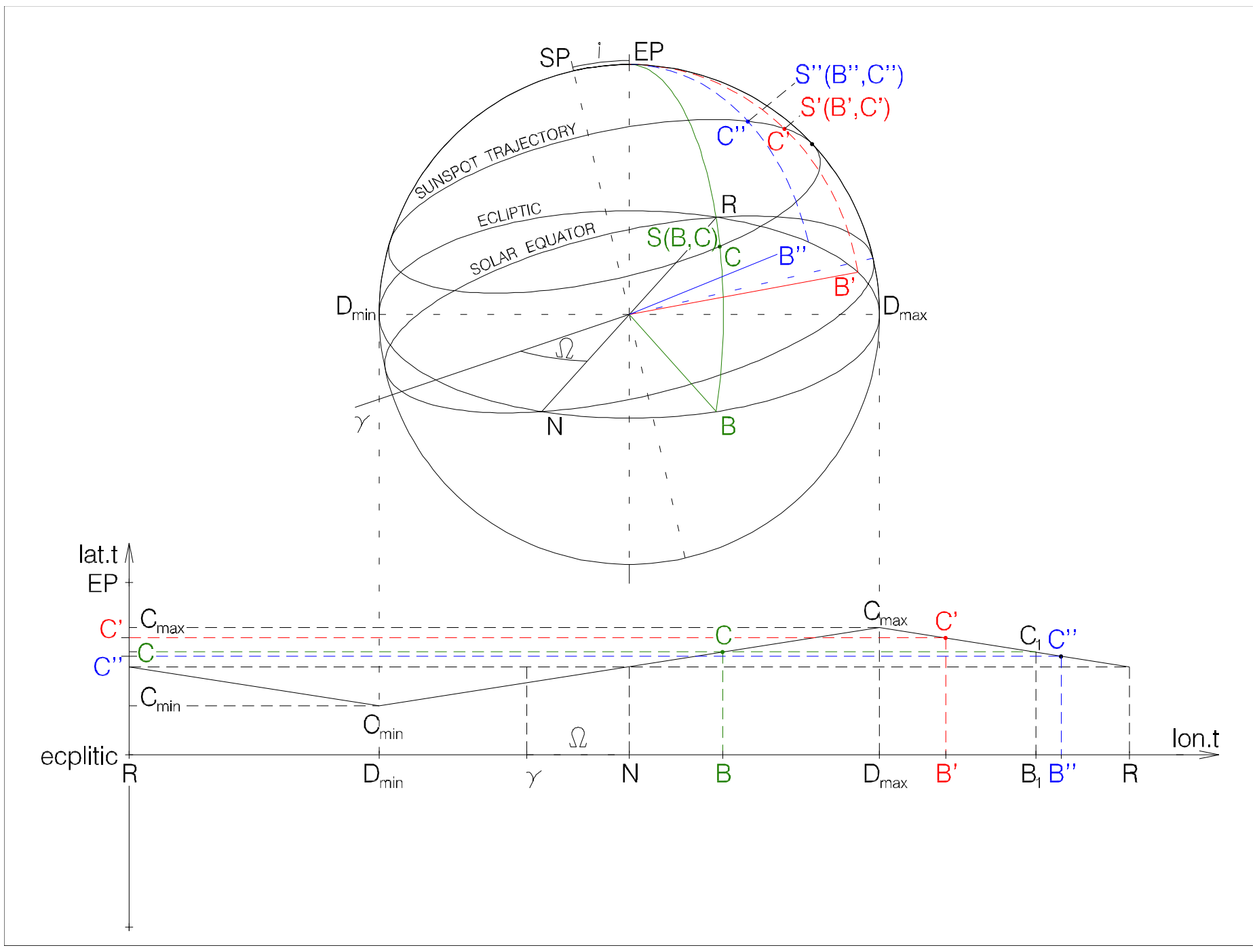

**Figure 1.** The method for Ω by Ruđer Bošković using: three positions of the same sunspot $S(B,C)$, $S'(B',C')$, and $S''(B'',C'')$ in the ecliptic coordinate system, $N = \Omega$ the ecliptic longitude of the ascending node from the vernal equinox ♈, $D_{min}$ and $D_{max}$ ecliptic longitudes of minimal and maximal elevation of the sunspot from the ecliptic.

Graphical representation of the method for Ω by Ruđer Bošković using three positions of the same sunspot is given in Figure 2. Bošković calculates $\Omega = N$ with a single equation where $f_N$ is a function of three sunspot positions $S(B,C)$, $S'(B',C')$, and $S''(B'',C'')$.

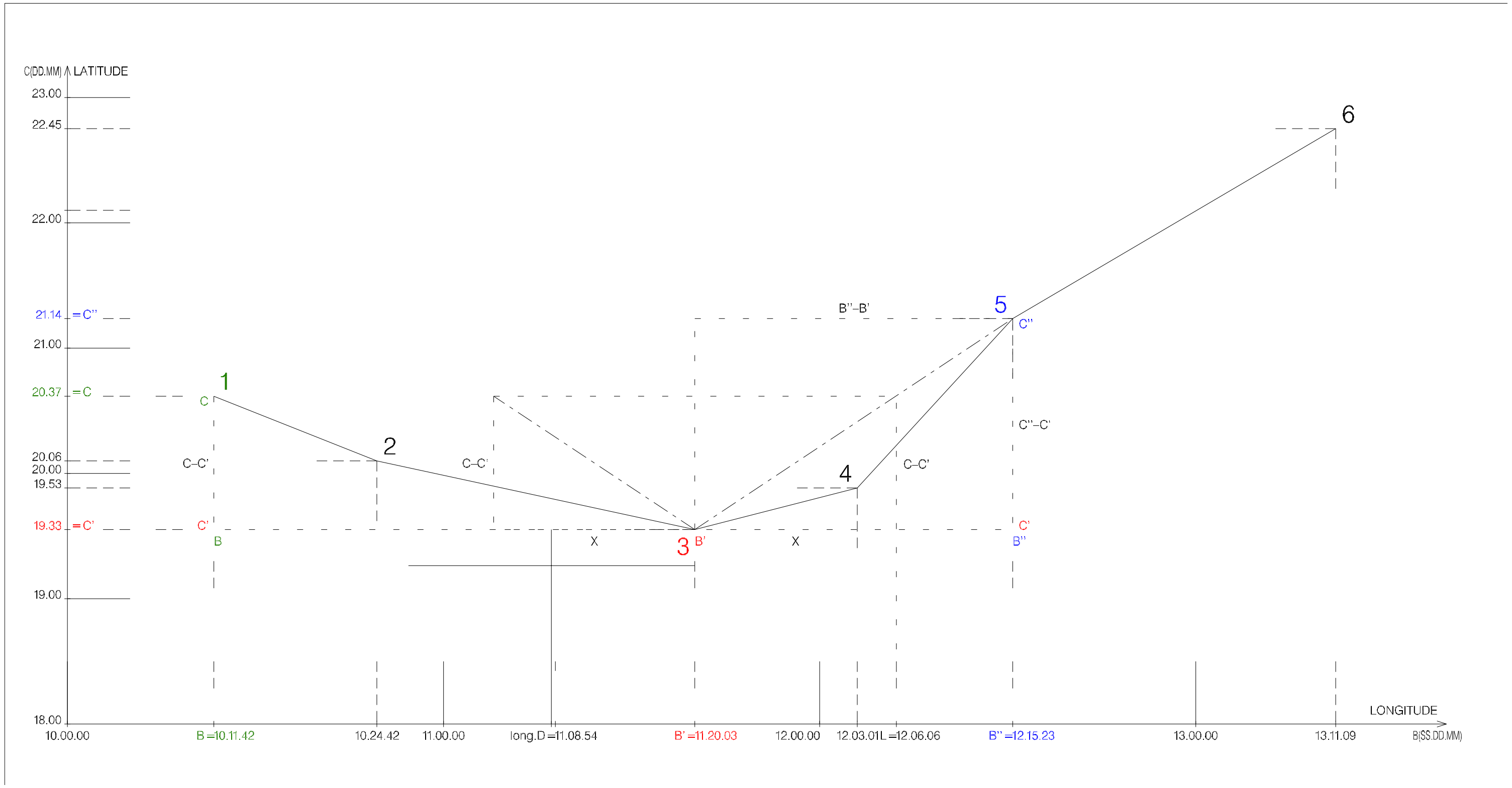

**Figure 2.**: Graphical representation of the method for determination of $\Omega$ by Ruđer Bošković using three positions of the same sunspot $S(B,C)$, $S'(B',C')$, and $S''(B'',C'')$ described in *Opuscule II* (Boscovich, 1785b, §.IV., №45).

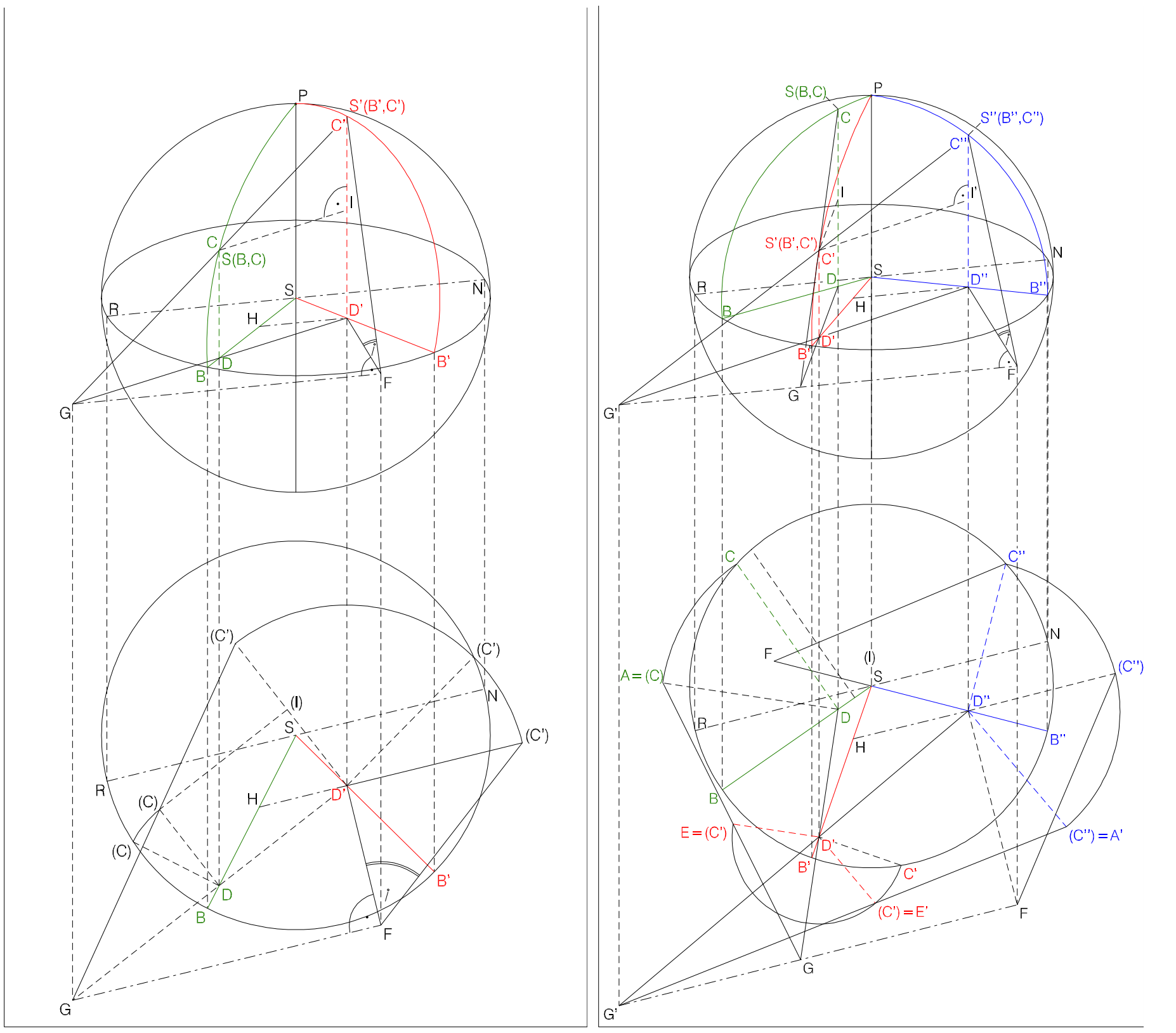

**Figure 3.** Axonometry (above) mapped to the ecliptic plane (below). (Left) The method for determination of $i$ by Ruđer Bošković using two positions of the sunspot $S(B,C)$, and $S'(B',C')$ and the ecliptic longitude of the node already known $N = \Omega$; (Right) The method for $\Omega$ and $i$ by Ruđer Bošković using three positions of the sunspot $S(B,C)$, $S'(B',C')$, and $S''(B'',C'')$.

## 2.2. The method for $i$

Bošković determined the solar equator inclination with planar trigonometric method using two sunspot positions with the ecliptic longitude of the ascending node already known. The method uses a planar triangle which is solved in two ways: using the planar graphical construction and the planar trigonometry solution (Boscovich, 1785b, §.V., №53), namely

$$i = f_i(N, B, B', C, C'). \tag{2}$$

In the present work, the planar trigonometrical method was used (Equation 2, Figure 3, left). Bošković calculates $i$ with a set of equations symbolically assigned with $f_i$ as a function of two sunspot positions $S(B,C)$, $S'(B',C')$, and previously calculated $\Omega = N$.

### 2.3. The method for periods, $T'$ and $T''$

The method determines the solar rotation period from two sunspot positions and already known elements, namely, the ecliptic longitude of the ascending node $\Omega$ and the solar equator inclination $i$. The method determines the angle in the pole of equatorial coordinate system between two declination arcs from two sunspot positions to the pole in the equatorial coordinate system (Equations 3 and 4). The period of rotation is calculated from the angular velocity (Boscovich, 1785b, §.VI.):

$$CP'D = f_{CP'D}(N, i, B, C), \tag{3}$$

$$T' = f_{T'}(CP'D, T.M._1, T.M._2). \tag{4}$$

Bošković calculates $T'$ with a set of equations symbolically showing $f_{CP'D}$ as a function of $N = \Omega$, $i$, and each sunspot position $S(B, C)$ of the same sunspot, and $f_{T'}$ as a function of the angle $CP'D$ and mean solar times $T.M._1$ and $T.M._2$.

Bošković calculated the synodic rotation period $T''$ using the arithmetic mean of six values of sidereal rotation period $\overline{T'}$ and duration of a year, $A = 365.25$ days (Equation 5):

$$T'' = \frac{A \cdot \overline{T'}}{A - \overline{T'}}. \tag{5}$$

### 2.4. The method for determination of $\Omega$ and $i$

The method determines the ecliptic longitude of the ascending node $\Omega = N$ and the solar equator inclination $i$ from three positions of the same sunspot (Boscovich, 1785b, §.VII.). Bošković presented three approaches:

1. The graphical method (Boscovich, 1785b, №70),
2. The planar trigonometry method (Boscovich, 1785b, №69), and
3. The spherical trigonometry method (Boscovich, 1785b, №76-№79).

In the present work the planar trigonometrical method was used (Equations 6 and 7, Figure 3, right):

$$\Omega = N = f_{N_3}(B, B', B'', C, C', C''), \tag{6}$$

$$i = f_{i_3}(B, B', B'', C, C', C''). \tag{7}$$

Bošković calculates $\Omega = N$ and $i$ with a set of equations symbolically involving $f_{N_3}$ and $f_{i_3}$ as functions of three sunspot positions $S(B, C)$, $S'(B', C')$, and $S''(B'', C'')$.

### 2.5. Bošković's methodology

Bošković's obtained several different results for each solar element, depending on which sunspot pairs or triplets he used. To combine those results, Bošković devised a methodology which consists of calculating arithmetic means of several values depending on the applied method, as follows (Boscovich, 1785b):

1. The ecliptic longitude of the ascending node $\Omega{=}N$ which is determined by the method for $\Omega{=}N$ in the *Tab. III.* (Boscovich, 1785b, *Tab. III.*, p. 167). It is the arithmetic mean of six, eight and ten pairs of the positions of the same sunspot, which Bošković used in the *Tab. IV.* (Boscovich, 1785b, *Tab. IV.*, p. 167).
2. The solar equator's inclination to the ecliptic which is determined by the method for $i$ in *Tab. V.* (Boscovich, 1785b, *Tab. V.*, p. 168). It is the arithmetic mean of five pairs of positions of the same sunspot, which Bošković used in the *Tab. VI.* (Boscovich, 1785b, *Tab. VI.*, p. 168).
3. Sidereal period $T'$, which is determined by the method for the period. It is the arithmetic mean of the six pairs of mean solar time $T.M.$ and the positions of the same sunspot, which Bošković used in the *Tab. X.* (Boscovich, 1785b, *Tab. X.*, p. 168).
4. The ecliptic longitude of the ascending node $N{=}\Omega$ and the solar equator inclination towards the ecliptic $i$, which were determined by the method for $\Omega$ and $i$. It uses three positions of the same sunspot in the *Tab. XII.* (Boscovich, 1785b, *Tab. XII.*, p. 169).

## 2.6. Methods for determination of the solar differential rotation

Since Bošković observed several sunspots at different heliographic latitudes, it is possible to investigate the variation of the rotation period (or angular velocity) with latitude, i.e., the differential rotation. Input data for determination of solar differential rotation are the sunspot positions in the ecliptic coordinate system and mean solar time. The sunspot positions were transformed into the heliographic coordinate system ($b$, $l$) using Equations 12 and 13. In order to calculate the synodic angular rotation rate $\omega_{syn}$ and solar differential rotation, for solar rotation parameters' $A$ and $B$ determination, we usually use two methods (Poljančić Beljan et al., 2014, equations 1 and 2 therein):

1. The daily shift (DS) method, which calculates angular rotation velocities from the daily differences of the central meridian distance (CMD) and the elapsed time or simply:

$$\omega_{syn} = \Delta CMD/\Delta t, \tag{8}$$

$$\omega_{sid} = \Delta l/\Delta T.M., \tag{9}$$

   where: $\omega_{syn}$ and $\omega_{sid}$ are the sunspot angular velocities, synodic and sidereal, $\Delta l$ is the difference of the heliographic longitudes, $\Delta T.M.$ is the difference of the mean solar times.
2. The linear least-squares fit method (LSQ) from the function *CMD(T.M.)* for each tracer or the function of sunspot positions. The LSQ method we use for:(i) fitting positions of tracers as function of time and calculating particular angular velocities and (ii) fitting the equation for the solar differential rotation law:

$$\omega(b) = A + B \cdot \sin^2 b. \tag{10}$$

In the present work, we calculated the angular velocity $\omega_{sid}$ by applying the daily shift method on the pairs of the same sunspot (Equation 9) and then the standard deviation for each sunspot

$$\sigma_{\omega_{sid}} = \sqrt{\frac{\sum (\Delta\omega_{sid})^2}{n-1}}, \tag{11}$$

where $\sum (\Delta\omega_{sid})^2$ is the sum of the squares of the deviations of arithmetic mean and $n$ is the number of determined angular velocities for the sunspot.

## 3. Results

In this work we present results which include the determination of sunspot positions and the elements of solar rotation. The latter was obtained using Bošković's observations of the first sunspot (repeating and modernizing his calculations), as well as all other Bošković's observations from 1777, as listed in the appendix of the *Opuscule II* (Boscovich, 1785b, *Appendice*).

### 3.1. The sunspot position determinations

The methods for the determination of the solar rotation elements use the positions of the same sunspot. The results are: original positions (Boscovich, 1785b, *Tab. II.*), repeated positions (Husak, Brajša, and Špoljarić, 2021a, Table 23) and repeated positions using corrected measurements[3] determined with streamlined equations for modern computers using $\varepsilon$ and $\varepsilon_{corr}$ (Table 1), and determination using corrected observations of four sunspots presented in Figure 4. Bošković used $lat.B.t = lat.t$ for ecliptic latitude in Boscovich (1785b).

### 3.2. Solar rotation elements' $\Omega$, $i$, $T'$, and $T''$ determinations

Ruđer Bošković determined the solar rotation elements[4] using his own methods (Boscovich, 1785b) in the present work shown in Table 2, row 1. We repeated

[3]Husak, Brajša, and Špoljarić (2021a) found several typing and other errors in Bošković (Boscovich, 1785b, *Appendice*).

[4]In the present work, we named solar rotation elements of original Bošković's example and present work (repeated) results as follows:

$\Omega_6$, $\Omega_8$ and $\Omega_{10}$ are the arithmetic means of ecliptic longitudes of the ascending node using six, eight and ten values (Boscovich, 1785b, *Tab. III.* and *Tab. IV.*);
$\Omega_{136}$ is the ecliptic longitude of the ascending node using three positions of the same sunspot (Boscovich, 1785b, *Tab. XII.*);
$i_5$ is the arithmetic mean of solar equator inclination using five values (Boscovich, 1785b, *Tab. V.* and *Tab. VI.*);
$i_{136}$ is the solar equator inclination using three positions of the same sunspot (Boscovich, 1785b, *Tab. XII.*);
$T'$ is the arithmetic mean of six values for sidereal solar rotation period (Boscovich, 1785b, *Tab. IX.* and *Tab. X.*).
$T''$ is the synodic solar rotation period (Boscovich, 1785b, *Tab. XI.*).

**Table 1.** The mean solar time and the ecliptic coordinates (*T.M.*, *lon.t*, *lat.t*) of the first sunspot: original positions (Boscovich, 1785b, *Tab. II.*), recalculated positions (Husak, Brajša, and Špoljarić, 2021a, Table 23, and recalculated positions using corrected measurements determined with streamlined equations for modern computers using the obliquity of the ecliptic $\varepsilon = 23°28'$ and $\varepsilon_{corr} = 23°27'47.1"$, and determination using corrected observations of four sunspots.

| Position | $\varepsilon = 23°28'$ (Boscovich, 1785b, *Tab. II.*) | | | $\varepsilon = 23°28'$ (Husak, Brajša, and Špoljarić, 2021a) | | |
|---|---|---|---|---|---|---|
| $N$ | $T.M.$[days] | $lon.t$[°] | $lat.B.t$[°] | $T.M.$[days] | $lon.t$[°] | $lat.B.t$[°] |
| 1 | 12.1257 | 311.7000 | 20.6167 | 12.1278 | 311.7000 | 20.6167 |
| 2 | 13.1056 | 324.7000 | 20.1000 | 13.1076 | 324.7000 | 20.1000 |
| 3 | 15.1299 | 350.0500 | 19.5500 | 15.1306 | 350.0500 | 19.5500 |
| 4 | 16.1549 | 3.0167 | 19.8833 | 16.1528 | 363.1000 | 19.8500 |
| 5 | 17.1375 | 15.3833 | 21.2333 | 17.1375 | 375.3667 | 21.2167 |
| 6 | 19.1042 | 41.1500 | 22.7500 | 19.1049 | 401.0500 | 22.7333 |
| Position | $\varepsilon = 23°28'$ | | | $\varepsilon_{corr} = 23°27'47.1"$ | | |
| $N$ | $T.M.$[days] | $lon.t$[°] | $lat.B.t$[°] | $T.M.$[days] | $lon.t$[°] | $lat.B.t$[°] |
| 1 | 12.1277 | 311.7119 | 20.6353 | 12.1277 | 311.7108 | 20.6331 |
| 2 | 13.1073 | 324.7173 | 20.1284 | 13.1073 | 324.7161 | 20.1268 |
| 3 | 15.1303 | 350.0545 | 19.5639 | 15.1303 | 350.0532 | 19.5637 |
| 4 | 16.1529 | 363.0301 | 19.9148 | 16.1529 | 363.0288 | 19.9154 |
| 5 | 17.1378 | 375.3833 | 21.2522 | 17.1378 | 375.3820 | 21.2534 |
| 6 | 19.1051 | 401.1625 | 22.7694 | 19.1051 | 401.1614 | 22.7719 |
| Position | For four sunspots | | | | | |
| $N$ | $T.M.$[days] | $lon.t$[°] | $lat.B.t$[°] | | | |
| 1 | 12.1277 | 311.7066 | 20.6141 | | | |
| 2 | 13.1073 | 324.7114 | 20.1087 | | | |
| 3 | 15.1303 | 350.0476 | 19.5473 | | | |
| 4 | 16.1529 | 363.0229 | 19.9002 | | | |
| 5 | 17.1376 | 375.3756 | 21.2395 | | | |
| 6 | 19.1051 | 401.1531 | 22.7601 | | | |

his calculations using original Bošković's equations (Table 2, row 2), equations streamlined for modern computers two times using the original and corrected obliquities of the ecliptic: $\varepsilon = 23°28'$ and $\varepsilon_{corr} = 23°27'47.1"$ (Table 2, rows 3 and 4).

To calculate the periods, we first transformed the ecliptic coordinates $lon.t = \lambda$ and $lat.t = \beta$ into heliographic ones, $b$ and $l$, by using Equations 12 and 13 and known values for $i$ and $\Omega$ (Waldmeier, 1955, Equations 3.7 and 3.8):

$$\sin b = \cos i \cdot \sin \beta - \sin i \cdot \cos \beta \cdot \sin(\lambda - \Omega), \tag{12}$$

$$\tan l = \cos i \cdot \tan(\lambda - \Omega) + \frac{\sin i \cdot \tan \beta}{\cos(\lambda - \Omega)}. \tag{13}$$

In Table 4, the sidereal period with its standard deviation was determined using Bošković's $i = 7°44'$ and $N = \Omega = 70°21'$ and *Tab. II.* mean solar time *T.M.* and ecliptic coordinates $lon.t = \lambda$ and $lat.B.t = \beta$ (Boscovich, 1785b, *Tab. II.*, *Tab. IV.*, and *Tab. VI.*) transformed into heliographic coordinates $b$ and $l$ shown in Table 3. The resulting periods, the sidereal $T'$ and synodic $T''$

**Table 2.**: The solar rotation elements using Bošković's observations of the first sunspot: 1. Boscovich (1785b) in row 1, recalculations using Bošković's original equations in row 2 and streamlined equations with different $\varepsilon$ values (rows 3 and 4).

| | $\Omega_6$ | $\Omega_8$ | $\Omega_{10}$ | $\Omega_{136}$ | $i_5$ | $i_{136}$ | $T'$[days] | $T''$[days] | Note |
|---|---|---|---|---|---|---|---|---|---|
| 1 Boscovich (1785b) | 70°21' | 71°32' | 73°09' | 74°03' | 7°44' | 6°49' | 26.77 | 28.89 | |
| Repeated procedure: | | | | | | | | | T.M. from: |
| 2 Original equations: | | | | | | | 26.77 | 28.89 | Boscovich (1785b) |
| | 70°20'52" | 71°31'46" | 73°11'35" | 74°03' | 7°45' | 6°48' | 26.76 | 28.87 | |
| Modernized equations: | | | | | | | | | |
| 3 $\varepsilon = 23°28'$ | 70°22' | 71°33' | 73°05' | 74°27' | 7°45' | 6°53' | 26.76 | 28.87 | |
| 4 $\varepsilon_{corr} = 23°27'47.1"$ | 70°21' | 71°32' | 73°04' | 74°25' | 7°45' | 6°53' | 26.76 | 28.87 | |

are: $T'_B = (26.7667 \pm 0.1495)$ days and $T''_B = 28.89$ days (Boscovich, 1785b), $T'_W = (26.7670 \pm 0.1433)$ days, and $T''_W = 28.8837$ days using Equations 12 and 13 (Waldmeier, 1955, Equations 3.7 and 3.8).

The differences between Bošković's and the present work's results are as follows. The differences of the arithmetic means and the standard deviations are: $\Delta T' = T'_B - T'_W = -0.0003$ days $= -26$ seconds and $\Delta\sigma = \sigma_B - \sigma_W = 0.0062$ days $= 8^{\mathrm{m}}57^{\mathrm{s}} = 537$ seconds. The differences $\Delta T'$ and $\Delta\sigma$ are not significant. Synodic period by Bošković is $T''_B = 28.89$ days (Boscovich, 1785a, *Tab. XI.*) and that determined using Waldmeier's values of the sidereal period by Bošković's equations is $T''_W = 28.8837$ days. They have minimal differences $\Delta T'' = T''_B - T''_W = 28.8872 - 28.8837 = 0.0035$ days $= 5^{\mathrm{m}}05^{\mathrm{s}} = 305$ seconds. Relative errors are: for the sidereal period $\Delta R_{T'} = \Delta T'/T'_B = -26^{\mathrm{s}}/26.7667$ days $= -0.001\%$ and for the accompanied standard deviation $\Delta R_\sigma = \Delta\sigma/\sigma_B = 8^a\mathrm{m}57^{\mathrm{s}}/0.1495$ days $= 4.1\%$ and for the synodic period $\Delta R_{T''} = \Delta T''/T''_B = +5^{\mathrm{m}}05^{\mathrm{s}}/28.89$ days $= 0.012\%$.

### 3.3. Solar differential rotation

In the present work, we determined the sidereal angular rotation rate $\omega_{sid}$ using the daily shift (DS) method from combinations of the convenient sunspot position pairs (Figure 4). The angular velocities were calculated from daily differences between two positions of the same sunspot. Angular velocity $\omega_{sid}$ and its standard deviation $\sigma_{\omega_{sid}}$ were determined using the pairs of sunspot positions (Equation 9 and 11).

The sunspots ordered by ascending heliographic latitudes $b$ are presented in Table 5, together with differences of heliographic latitudes $\Delta b[°]$, angular velocities $\Delta\omega[°/\mathrm{day}]$, and periods $\Delta T[\mathrm{days}]$ and heliographic latitudes $b[°]$, angular velocities $\omega[°/\mathrm{day}]$ and periods $T'[\mathrm{days}]$.

Today we know that the solar rotation is differential. Bošković's 1777 sunspot observations (consequently $\omega_{sid}$, $\sigma_{\omega_{sid}}$, and $b$) show that with increasing heliographic latitudes $b$ the angular velocity $\omega$ decreases. The dependence of sunspot angular velocity $\omega_{sid}$ and heliographic latitude $b$, as measured from Bošković observations, is shown in Figure 5, and compared with the solar differential rotation profile determined by Ruždjak et al. (2017) from the Greenwich Photoheliographic Results sunspot dataset (1874–1976) (Ruždjak et al., 2017, Table 1: GPR dataset 1874-1976).

## 4. Analysis

In this section, we compare the values of solar rotation elements $\Omega$, $i$, and periods $T'$ and $T''$, obtained using various methods with Bošković's original results. Table 6 lists differences between Bošković's values and our recalculations using Bošković's original equations (first row), using modern equations with two different $\varepsilon$ values (second and third row), and finally using Carrington values precessed for the year 1777 (last row).

Bošković determined the ecliptic longitude of the ascending node using his two methods: arithmetic mean of six pairs of positions of the same sunspot $\Omega_6 =$ $70°21' = 70.35°$, and the method using three positions of the same sunspot $\Omega_{136} = 74°03' = 74.05°$. He determined the inclination of the solar equator towards the ecliptic using two methods: the arithmetic mean of five pairs using two positions of the same sunspot and the already known ecliptic longitude of the ascending node $i_5 = 7°44' = 7.73°$, and the method using three positions of the same sunspot $i_{136} = 6°49' = 6.82°$.

**Table 3.**: The observations of the first sunspot are from September 1-6, 1777: mean solar time *T.M.* ($j$ date in September 1777, where $j$ stands for day in French) and ecliptic coordinates $lon.t = \lambda$ and $lat.B.t = \beta$ in *Tab. II.*, Bošković's $N = \Omega = 70°21'$ in *Tab. IV.* and $i = 7°44'$ in *Tab. VI.* (Boscovich, 1785b, *Tab. II.*, *Tab. IV.* and *Tab. VI.*)) transformed into heliographic coordinates $l$ and $b$ (Waldmeier, 1955, equations 3.7 and 3.8) with supplements $\Delta\lambda$ for $\lambda$ and $\Delta l$ for $l$.

| | | | | | | |
|---|---|---|---|---|---|---|
| $\Omega =$ | 70° | 21' | *Tab. IV.* (Boscovich, 1785b) | $\Omega =$ | 1.227839 | $1° = \pi/180° = 0.017453$ rad |
| $i =$ | 7° | 44' | *Tab. VI.* (Boscovich, 1785b) | $i =$ | 0.134972 | $1$ rad $= 180°/\pi = 57.29578°$ |

*Tab. II.* (Boscovich, 1785b) — Heliographic coordinates $l$, $b$ calculated from Bošković's $\lambda$, $\beta$ using equations 3.7 and 3.8 (Waldmeier, 1955).

| | T.M. j | h | m | s | $lon.t = \lambda$ ° | ' | $lat.B.t = \beta$ ° | ' | $\Delta\lambda$ [°] | $T.M. + 0.5$ [days] | $\lambda$ [rad] | $\beta$ [rad] | $l$ [°] | $b$ [°] | $\Delta l$ [°] |
|---|---|---|---|---|---|---|---|---|---|---|---|---|---|---|---|
| 1 | 12 | 3 | 1 | 10 | 11 | 42 | 20 | 37 | | 12.6257 | 5.4402 | 0.3598 | 59.6533 | 27.3507 | |
| 2 | 13 | 2 | 32 | 10 | 24 | 42 | 20 | 6 | | 13.6056 | 5.6671 | 0.3508 | 73.4006 | 27.5302 | |
| 3 | 15 | 3 | 7 | 11 | 20 | 3 | 19 | 33 | | 15.6299 | 6.1095 | 0.3412 | 100.2803 | 27.1665 | 180 |
| 4 | 16 | 3 | 43 | 0 | 3 | 1 | 19 | 53 | 360 | 16.6549 | 6.3358 | 0.3470 | 113.9963 | 26.9865 | 180 |
| 5 | 17 | 3 | 18 | 0 | 15 | 23 | 21 | 14 | 360 | 17.6375 | 6.5517 | 0.3706 | 127.0981 | 27.4888 | 180 |
| 6 | 19 | 2 | 30 | 1 | 11 | 9 | 22 | 45 | 360 | 19.6042 | 7.0014 | 0.3971 | 153.9341 | 26.3424 | 180 |

**Table 4.**: Bošković's *Tab. X.* (Boscovich, 1785b): the sunspot position pairs and sidereal periods $T'$[days], heliographic longitudes $l$[°] and mean solar times $T.M.$[days] of pairs, differences of longitudes $\Delta l$[°] and mean solar times $\Delta T.M.$[days], angular velocity $\omega$[°/day], sidereal period $T'$[days], the sum and arithmetic mean with standard deviations and period differences in days $\Delta T'$[days] and seconds $\Delta T'$[s] determined using Boscovich (1785) and Waldmeier (1955) procedure.

| Boscovich (1785b) | | | Angular velocity $\omega$ and solar rotation period $T'$ determined using $l$, $b$ (Waldmeier, 1955). | | | | | | | | | | T(Bošković)-T(Waldmeier) | |
|---|---|---|---|---|---|---|---|---|---|---|---|---|---|---|
| *Tab. X.* | | $T'$[days] | $l_1$ | $l_2$ | $T.M._1$ | $T.M._2$ | $\Delta l$ | $\Delta T.M.$ | $\omega$[°/day] | $T'$[days] | $\Delta T'$ | $\Delta T'^2$ | $\Delta T'$[days] | $\Delta T'$[s] |
| 4 | 1 | 26.69 | 113.9963 | 59.6533 | 16.6549 | 12.6257 | 54.3430 | 4.0292 | 13.4874 | 26.6916 | -0.0754 | 0.0057 | -0.0016 | - 134.0580 |
| 5 | 1 | 26.75 | 127.0981 | 59.6533 | 17.6375 | 12.6257 | 67.4448 | 5.0118 | 13.4572 | 26.7515 | -0.0155 | 0.0002 | -0.0015 | -131.1009 |
| 6 | 1 | 26.63 | 153.9341 | 59.6533 | 19.6042 | 12.6257 | 94.2808 | 6.9785 | 13.5102 | 26.6465 | -0.1205 | 0.0145 | -0.0165 | -1421.6825 |
| 5 | 2 | 27.04 | 127.0981 | 73.4006 | 17.6375 | 13.6056 | 53.6975 | 4.0319 | 13.3180 | 27.0311 | 0.2641 | 0.0697 | 0.0089 | 771.3020 |
| 6 | 2 | 26.82 | 153.9341 | 73.4006 | 19.6042 | 13.6056 | 80.5335 | 5.9986 | 13.4254 | 26.8149 | 0.0479 | 0.0023 | 0.0051 | 438.9286 |
| 6 | 3 | 26.67 | 153.9341 | 100.2803 | 19.6042 | 15.6299 | 53.6538 | 3.9743 | 13.5002 | 26.6663 | -0.1007 | 0.0101 | 0.0037 | 318.8558 |
| | | 160.62 | [°] | [°] | [days] | [days] | [°] | [days] | $\Sigma =$ | 160.6018 | 0.0000 | 0.1026 | A [days]= | 365.25 |
| $T'_B =$ | | 26.77 | ± | 0.1495 | | | | | $T'_W =$ | 26.7670 | ± | 0.1433 | $T''_W =$ | 28.8837 |

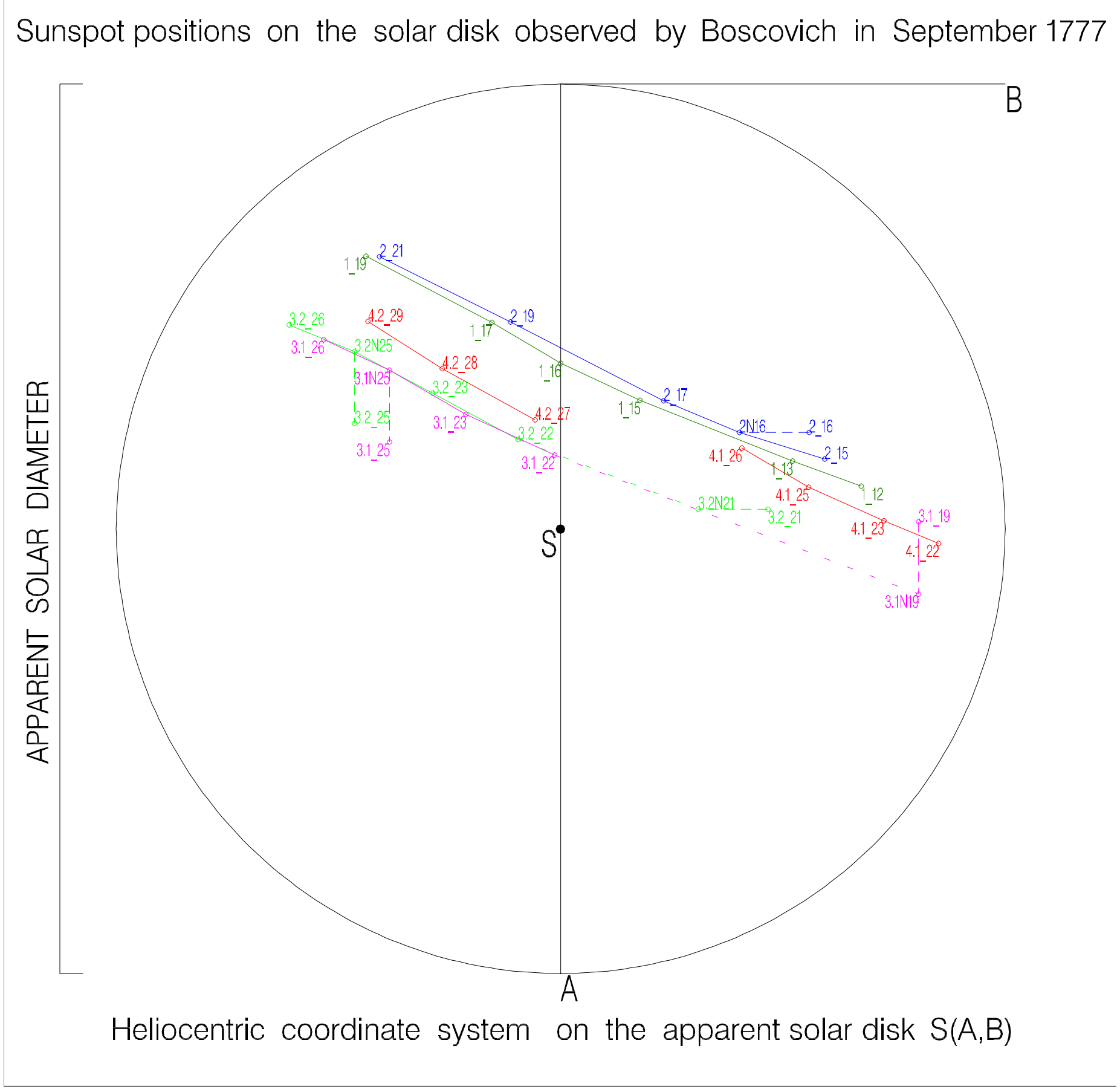

**Figure 4.** The sunspot observations of Bošković in September 1777 mapped on the apparent solar disk, gross errors are for: sunspot 2 (September 16th), sunspot 3.1 (September 19th and 25th) and sunspot 3.2 (September 21st and 25th). Gross-errors present vectors from the position using Bošković's observation measurements (_) and corrected (new) position along the expected trajectory (N) of the sunspot: (right-to-left): 2_16 – 2N16 (sunspot 2, 16 September, 1777) and 3.2_21 – 3.2N21 (sunspot 3.2, 21 September, 1777); (down-up): 3.1_19 – 3.1N19 (sunspot 3.1, 19 September, 1777), 3.1_25 – 3.1N25 (sunspot 3.1, 25 September, 1777) and 3.2_25 – 3.2N25 (sunspot 3.2, 25 September, 1777) (local coordinates are measured from solar disk center, with negative to the left and positive to the right (A), and from the upper border (B)).

In Table 7 and Figure 6, we compare Bošković results for $i$ and $\Omega$ with results obtained by other observers. The table is taken from Wöhl (1978), Table 1 and

**Table 5.** Solar sidereal rotation angular velocities and sidereal rotation periods using all Bošković's observations in September 1777 from *Appendice* of *Opuscule II*: Heliographic latitudes $b[°]$ (south-to-north), angular velocities $\omega[°/\text{day}]$ with their standard deviation $\sigma_\omega[°/\text{day}]$ and sidereal solar rotation periods $T'[\text{days}]$ on the basis of heliographic coordinates $b$ and $l$, Bošković's number of the sunspot, pairs(:)/positions were used for determination of heliographic latitude $b$, angular velocity $\omega$ and period $T$ of the sunspots 1, 2, 3, 4.1, and 4.2. The lower part of the table presents the differences $\Delta b[°]$, $\Delta\omega[°/\text{day}]$ and $\Delta T[\text{day}]$ of sunspots, positions used for calculating the arithmetic means of $b[°]$ and calculated $\omega[°/\text{day}]$ and $T'[\text{days}]$.

Heliographic latitude $b$, angular velocity $\omega$ and period $T$ of the sunspots 1, 2, 3, 4.1 i 4.2. ordered from south to north.

| Sunspot | Pairs(:)/position | $b[°]$ | $\omega[°/\text{day}]$ | $\pm\sigma_\omega[°/\text{day}]$ | $T'[\text{days}]$ |
|---|---|---|---|---|---|
| 3 | 3:5 and 4:6 | 15.054275 | 14.4045 | 0.2531 | 24.9960 |
| 4.2 | 1, 2, 3, 4 | 18.034242 | 15.5776 | 0.3531 | 23.1180 |
| 4.1 | 5, 6, 7 | 25.276158 | 11.0043 | 2.1053 | 33.7072 |
| 1 | ALL | 27.150421 | 13.4756 | 0.1872 | 26.7196 |
| 2 | ALL except 2nd position* | 28.358973 | 13.9618 | 0.3657 | 25.7995 |

| $\Delta b[°]$ | $\Delta\omega[°/\text{day}]$ | $\Delta T[\text{days}]$ | Sunspot | Positions | $b[°]$ | $\omega[°/\text{day}]$ | $T'[\text{days}]$ |
|---|---|---|---|---|---|---|---|
| - | - | - | 3 | 3:5 and 4:6 | 15.054275 | 14.4045 | 24.9960 |
| 2.98 | 1.1731 | −1.8780 | 4.2 | 1, 2, 3, 4 | 18.034242 | 15.5776 | 23.1180 |
| 7.24 | −4.5733 | 10.5892 | 4.1 | 5, 6, 7 | 25.276158 | 11.0043 | 33.7072 |
| 1.87 | 2.4713 | -6.9876 | 1 | all | 27.150421 | 13.4756 | 26.7196 |
| 1.99 | 1.1836 | 2.6782 | 2 | all | 29.136168 | 14.6592 | 29.3978 |

*sunspot 2 has gross-error for 2nd position on 16 September, 1777 (vector 2_16-2N16, Figure 4).

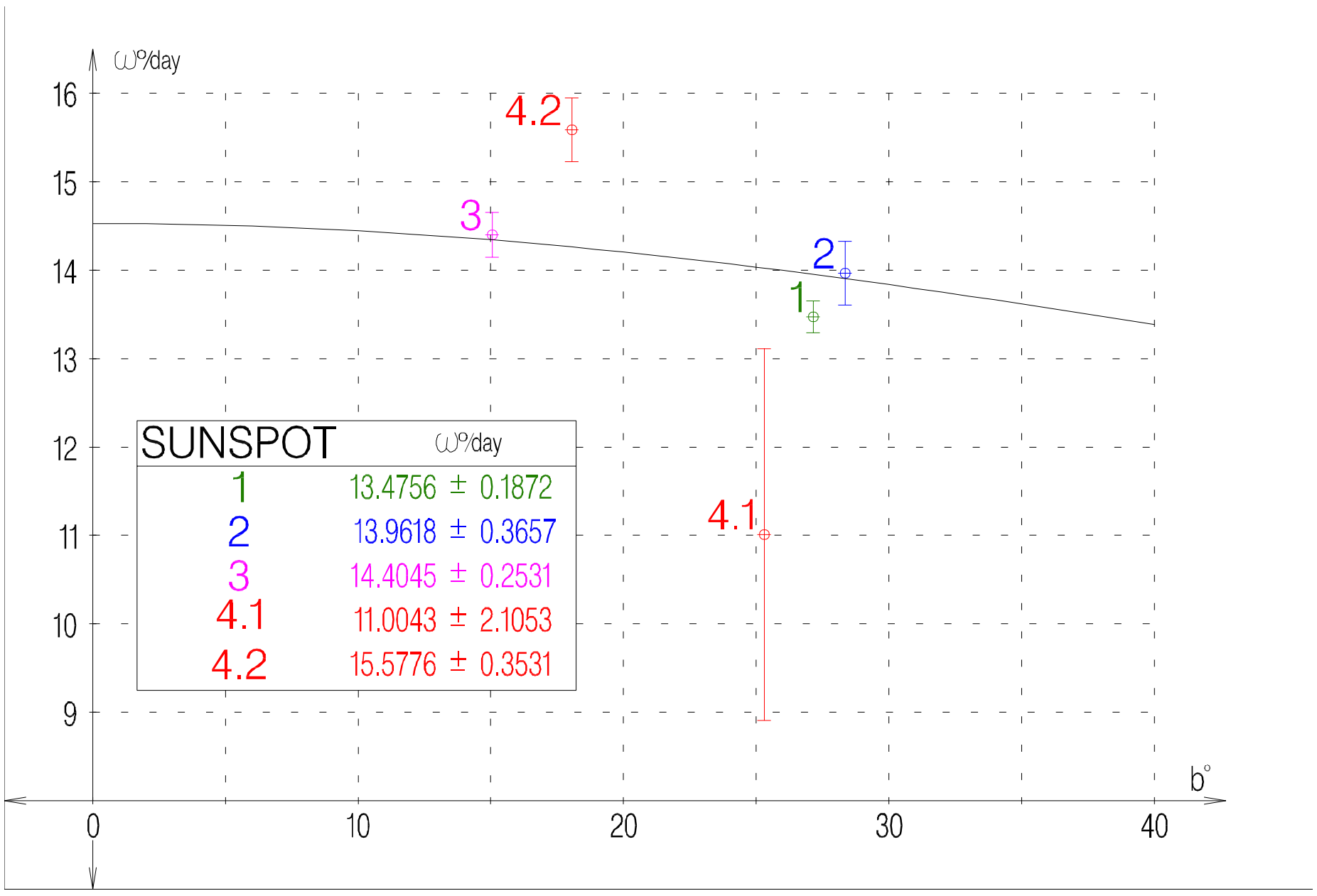

**Figure 5.** Solar sidereal rotation angular velocities calculated from observations of Bošković, data taken from the upper part of Table 5 with determinations of angular velocities $\omega_{sid}$, standard deviations $\sigma_{\omega_{sid}}$ and heliographic latitudes $b$. For comparison, differential rotation profile of sunspot groups in the period 1874–1976 $\omega(b) = A + B \cdot \sin^2 b$, $A = (14.528 \pm 0.006)[° \cdot \text{day}^{-1}]$, $B = (-2.77 \pm 0.05)[° \cdot \text{day}^{-1}]$ is given (Ruždjak et al., 2017, Table 1: GPR dataset 1874-1976).

**Table 6.**: The differences of the solar rotation elements $\Omega$, $i$, $T'$, $T''$, between the values of original Bošković's example (row 1 in Table 2) and recalculated values (rows 2, 3 and 4 in Table 2), the last row gives the differences between the values in row 1 of Table 2 from Carrington's values $\Omega_{C_{1777}} = 72.6476°$ and $i_C = 7.25°$.

| Procedure | $\varepsilon$ | $\Delta\Omega_6$ | $\Delta\Omega_8$ | $\Delta\Omega_{10}$ | $\Delta\Omega_{136}$ | $\Delta i_5$ | $\Delta i_{136}$ | $\Delta T'$ | $\Delta T''$ |
|---|---|---|---|---|---|---|---|---|---|
| | ° ' " | ' | ' | ' | ' | ' | ' | days | days |
| Original for the first sunspot | $\varepsilon = 23°28'$ | 0 | 0 | −3 | 0 | 1 | −1 | 0.00 | 0.00 |
| Modified equations for the first sunspot | $\varepsilon = 23°28'$ | 1 | 1 | −4 | 24 | 1 | 4 | −0.01 | −0.02 |
| Modified equations for the first sunspot | $\varepsilon_{corr} = 23°27'47.1''$ | 0 | 0 | −5 | 22 | 1 | 4 | −0.01 | −0.02 |
| Modified equations for four sunspots | $\varepsilon_{corr} = 23°27'47.1''$ | −2°26' | −1°20' | 0°30' | 1°24' | 29' | −26' | - | - |

supplemented with verified results from Bošković. Due to precession, longitude of the ascending node had to be recalculated to a specific year for accurate comparison. We used the year of observation and years 1900 and 1976, as in Wöhl (1978). Precession-corrected $\Omega$ was calculated using Equation 14, while the relative error to Carrington values was calculated using Equation 15:

$$\Omega_Y = \Omega_{Obs} + (Y - Y_{Obs}) \cdot 0.01396°, \tag{14}$$

$$R_\Omega = \frac{\Omega_{Carr} - \Omega_Y}{\Omega_{Carr}} \cdot 100\%, \tag{15}$$

where $Y$ and $Y_{Obs}$ represent the selected year and year of observation, respectively.

We recalculated the ecliptic longitudes of the ascending node $\Omega_{1900}$ and $\Omega_{1976}$ for the years 1900 and 1976 using Equation 14. Then we calculated their differences from the values of year of observation and $\Omega_{Obs}$ by Wöhl, 1978, Table 1.

The values $\Delta\Omega_{1900}^{W-PW}$ and $\Delta\Omega_{1976}^{W-PW}$ are slightly different from the present work values $\Omega_{1900}^{PW}$ and $\Omega_{1976}^{PW}$ and ecliptic longitudes of the ascending node $\Omega_{1900} = \Omega_{1900}^{W}$ and $\Omega_{1976} = \Omega_{1976}^{W}$ given by Wöhl (1978).

Relative errors were determined using the differences of the Carrington's $\Omega_{1900}^{C} = 74.36°$ and $\Omega_{1976}^{C} = 75.43°$ for the certain year as reference values and present work $\Omega_{1900}^{PW}$ and $\Omega_{1976}^{PW}$. In Table 7 the relative errors for $\Omega_6$ from pairs and $\Omega_{136}$ from triples determined by Bošković regarding the Carrington's values $\Omega_{1900}^{C} = 74.36°$ and $\Omega_{1976}^{C} = 75.43°$ for the certain year were calculated (Equation 15): from $\Omega$'s from pairs ($\Omega_{6(1900)} = 72.07°$ and $\Omega_{6(1976)} = 73.13°$) $R_{\Omega_{(1900)}} = 3.09\%$ and $R_{\Omega_{(1976)}}^{C-PW} = 3.05\%$ and $\Omega$'s from triples ($\Omega_{136(1900)} = 75.77°$ and $\Omega_{136(1976)} = 76.83°$) $R_{\Omega_{136(1900)}}^{C-PW} = -1.89\%$ and $R_{\Omega_{136(1976)}} = -1.86\%$ (Table 7).

The relative errors $R_{\Omega_{1900}}^{C-PW} \approx R_{\Omega_{1976}}^{C-PW}$ are $-13.67\% < R_{\Omega_{1900}}^{C-PW} < 4.39\%$ and $-13.48\% < R_{\Omega_{1976}}^{C-PW} < 4.33\%$. Among other results, Bošković's results are very good, the relative errors being from pairs $R_{\Omega_{6(1900)}}^{C-PW} \approx R_{\Omega_{6(1976)}}^{C-PW} \approx 3.10\%$ and from triples $R_{\Omega_{136(1900)}}^{C-PW} \approx R_{\Omega_{136(1976)}}^{C-PW} \approx -1.90\%$, so their absolute values are approximately 3% and 2%, respectively.

We can see from Table 7 that the relative errors of various observers in comparison with Carrington values are between 4% and −14%, while Bošković's values show a really good agreement with relative error values below 3%. This confirms the validity and precision of his observation and calculation methods.

Bošković discussed the ecliptic longitude of the ascending node $\Omega = N$, the solar equator inclination $i$, and the solar rotation period $T'$ which are taken from his discussion (Boscovich, 1785b, *§.XIV. Réflexions*). The discussion includes reflexions (*§.XIV. Réflexions*) of Ruđer Bošković in *Opuscule II* (Boscovich, 1785b). He discussed obtained values regarding the known results[5] presented in Table 8.

[5]In Bošković's time, the Zodiac sign $1^{s} = 30°$ usually was used.

**Table 7.**: The solar rotation elements $\Omega$ and $i$ as derived from measurements of the rotation of sunspots (all angles in degrees) (Wöhl, 1978, Table 1) supplemented with Bošković's 1777 results (underlined) and $\Omega$ and $i$ determinations for the last hundred years (Wöhl, 1978*; Stark and Woehl, 1981**; Balthasar et al., 1986***; Balthasar, Woehl, and Stark, 1987****; Beck and Giles, 2005*****), presented in Figure 6.

| Observer | Year of obs. | $\Omega_{obs}$ | $\Omega_{1900}$ | $\Omega_{1976}$ | $i$ | Source: | Present work | | $\Omega_Y^W - \Omega_Y^{PW}$ | | $\Omega_Y^C - \Omega_Y^{PW}$ | | Relative errors [%] | | $\Omega^C - \Omega^W$ | |
|---|---|---|---|---|---|---|---|---|---|---|---|---|---|---|---|---|
| | | | | | | | $\Omega_{1900}^{PW}$ | $\Omega_{1976}^{PW}$ | $\Delta\Omega_{1900}^{W-PW}$ | $\Delta\Omega_{1976}^{W-PW}$ | $\Delta\Omega_{1900}^{C-PW}$ | $\Delta\Omega_{1976}^{C-PW}$ | $R_{\Omega_{1900}}^{C-PW}$ | $R_{\Omega_{1976}}^{C-PW}$ | $\Delta\Omega_{1900}^{C-W}$ | $\Delta\Omega_{1976}^{C-W}$ |
| J. D. Cassini | 1678 | 68 | 71.1 | 72.2 | 7.5 | (Wöhl, 1978, Table 1) | 71.10 | 72.16 | 0.00 | 0.04 | 3.27 | 3.27 | 4.39% | 4.33% | 3.26 | 3.23 |
| J. Cassini | 1740 | 70 | 72.2 | 73.3 | - | (Wöhl, 1978, Table 1) | 72.23 | 73.29 | -0.03 | 0.01 | 2.13 | 2.13 | 2.87% | 2.83% | 2.16 | 2.13 |
| La Lande | 1775 | 78.07 | 79.81 | 80.87 | 7.33 | (Wöhl, 1978, Table 1) | 79.82 | 80.88 | 0.00 | -0.01 | -5.45 | -5.45 | -7.33% | -7.23% | -5.45 | -5.44 |
| Delambre | 1775 | 80.12 | 81.86 | 82.92 | 7.32 | (Wöhl, 1978, Table 1) | 81.87 | 82.93 | -0.01 | -0.01 | -7.50 | -7.50 | -10.09% | -9.94% | -7.50 | -7.49 |
| Ruđer Bošković[2] | 1777 | <u>70.35</u> | <u>72.07</u> | <u>73.13</u> | <u>7.73</u> | Present work (from pairs)[2] | <u>72.07</u> | <u>73.13</u> | – | – | <u>2.30</u> | <u>2.30</u> | <u>3.09%</u> | <u>3.05%</u> | – | – |
| Ruđer Bošković[3] | 1777 | <u>74.05</u> | <u>75.77</u> | <u>76.83</u> | <u>6.82</u> | Present work (from triples)[3] | <u>75.77</u> | <u>76.83</u> | – | – | <u>-1.40</u> | <u>-1.40</u> | <u>-1.89%</u> | <u>-1.86%</u> | – | – |
| Böhm | 1833 | 76.78 | 77.72 | 78.78 | 6.95 | (Wöhl, 1978, Table 1) | 77.72 | 78.78 | 0.00 | 0.00 | -3.35 | -3.35 | -4.51% | -4.44% | -3.36 | -3.35 |
| Laugier | 1840 | 75.13 | 75.97 | 77.03 | 7.15 | (Wöhl, 1978, Table 1) | 75.97 | 77.03 | 0.00 | 0.00 | -1.60 | -1.60 | -2.16% | -2.13% | -1.61 | -1.60 |
| Wiehmann | 1846 | 83.78 | 84.53 | 85.59 | 7.75 | (Wöhl, 1978, Table 1) | 84.53 | 85.59 | 0.00 | 0.00 | -10.17 | -10.17 | -13.67% | -13.48% | -10.17 | -10.16 |
| Carrington | 1850 | 73.67 | 74.37 | 75.43 | 7.25 | (Wöhl, 1978, Table 1) | 74.37 | 75.43 | 0.00 | 0.00 | 0.00 | 0.00 | 0.00% | 0.00% | -0.01 | 0.00 |
| Spoerer | 1861/66 | 74.52 | 75.07 | 76.13 | 6.97 | (Wöhl, 1978, Table 1) | 75.06 | 76.13 | 0.01 | 0.00 | -0.70 | -0.70 | -0.94% | -0.93% | -0.71 | -0.70 |
| Wilsing | 1882 | 75.78 | 76.03 | 77.09 | 7.17 | (Wöhl, 1978, Table 1) | 76.03 | 77.09 | 0.00 | 0.00 | -1.67 | -1.67 | -2.24% | -2.21% | -1.67 | -1.66 |
| (Data above – except column 5 – as cited by Epstein, 1904) | | | | | | | | | | | | | | | | |
| Dyson and Maunder (1912) | 1874–1911 | – | 74.59 | 75.65 | 7.18 | (Wöhl, 1978, Table 1) | – | – | – | – | – | – | – | – | -0.23 | -1.22 |
| Dyson and Maunder (1913) | 1874–1912 | – | 74.48 | 75.54 | 7.18 | (Wöhl, 1978, Table 1) | – | – | – | – | – | – | – | – | -0.12 | -0.11 |
| Epstein (1916) | 1905–1910 | – | 76.30 | 77.36 | 7.20 | (Wöhl, 1978, Table 1) | – | – | – | – | – | – | – | – | -1.94 | -1.93 |
| Epstein (1917) | 1903–1910 | – | 73.97 | 75.03 | 7.27 | (Wöhl, 1978, Table 1) | – | – | – | – | – | – | – | – | 0.39 | 0.40 |
| H. Wöhl* | 1976 | $76.31 \pm 0.65$ | – | – | $6.77 \pm 0.31$ | (Wöhl, 1978) | – | – | – | – | – | – | – | – | – | – |
| D. Stark and H. Wöhl** | 1980 | $75.8 \pm 0.27$ | – | – | $7.15 \pm 0.034$ | (Stark and Woehl, 1981) | – | – | – | – | – | – | – | – | – | – |
| Balthasar et al.*** | 1874–1984 | $73.75 \pm 0.15$ | – | – | $7.137 \pm 0.017$ | (Balthasar et al., 1986) | – | – | – | – | – | – | – | – | – | – |
| Balthasar, Stark, Wöhl**** | 1874–1976 | $73.86 \pm 0.38$ | – | – | $7.12 \pm 0.05$ | (Balthasar, Woehl, and Stark, 1987) | – | – | – | – | – | – | – | – | – | – |
| J. G. Beck and P. Giles***** | 1996–2001 | $73.5 \pm 0.1$ | – | – | $7.155 \pm 0.002$ | (Beck and Giles, 2005) | – | – | – | – | – | – | – | – | – | – |

($\Omega$ increases by about 0.01396°/year by the precesion.)

Note: $\Omega_Y^{Present\ work} = \Omega_Y^{PW}$, $\Omega_Y^{Woehl(1978)} = \Omega_Y^W$, $\Omega_Y^{Carrington} = \Omega_Y^C$ for the years $Y = 1900$ and $Y = 1976$ are $\Omega_{1900}^C = 74.36°$ and $\Omega_{1976}^C = 75.43°$.

*Doppler Velocity Measurements of the Solar Plasma (Wöhl, 1978).

**The solar rotation elements determined from sunspot observations in the years 1940 to 1968 given in the Greenwich Photoheliographic results (Stark and Woehl, 1981).

***The solar rotation elements derived from sunspot group observations obtained at Greenwich 1874 to 1986 and at Kanzelhöhe 1947 to 1984, determination of $\Omega_{1850}$ for the year 1850 (Balthasar et al., 1986).

****The solar rotation elements derived from recurrent single sunspots, determination of $\Omega_{1850}$ for the year 1850 (Balthasar, Woehl, and Stark, 1987).

*****Helioseismic Determination of the Solar Rotation Axis (Beck and Giles, 2005).

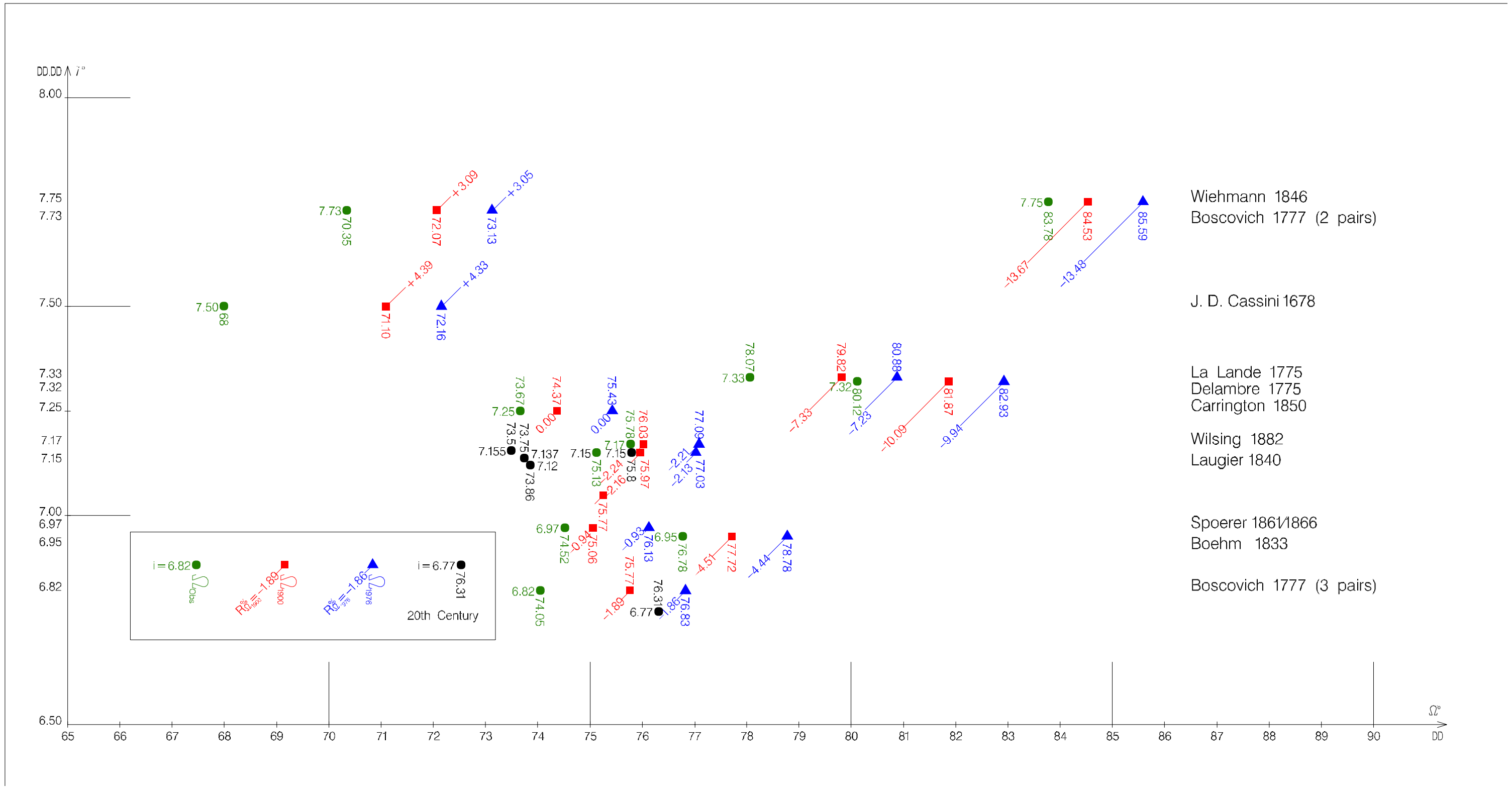

**Figure 6.**: The solar rotation elements $\Omega_{Obs}$, $\Omega_{1900}^{PW}$, $\Omega_{1976}^{PW}$ and $i$ as derived from measurements of the rotation of sunspots (all angles in degrees) (Wöhl, 1978, Table 1) supplemented with Bošković's 1777 results and $\Omega$ and $i$ determinations for the last hundred years (Wöhl, 1978*; Stark and Woehl, 1981**; Balthasar et al., 1986***; Balthasar, Woehl, and Stark, 1987****; Beck and Giles, 2005*****), presented in Table 7.

**Table 8.**: The solar rotation elements Ω, $i$ and sidereal period $T'$ determined by Bošković and by other researchers before him taken from *§.XIV. Réflexions...*, paragraphs numbered №141-№165 (Boscovich, 1785b, *§.XIV. Réflexions...*, №141-№165).

| Researcher | $N = \Omega[^{s}\ ^{\circ}\ ']$[a] | Note for $N$ | $i[^{\circ}]$ | Note for $i$ | $T$[days] | Note for $T$ |
|---|---|---|---|---|---|---|
| Ruđer Bošković | $2^{s}14^{\circ}03'$ | Using three sunspot positions in *Tab. XII.* | $6^{\circ}49'$ | Using three sunspot positions in *Tab. XII.* | $26\frac{3}{4}$ | $n = 6$ |
| Ruđer Bošković | $2^{s}10^{\circ}21'$ | $n = 6$ (without $14^{\circ}51'$ and $15^{\circ}18'$) | $7^{\circ}44'$ | Arithmetic mean of five values in *Tab. VI.* | | |
| Ruđer Bošković | $2^{s}11^{\circ}32'$ | $n = 8$ | | | | |
| Ruđer Bošković | $2^{s}13^{\circ}09'$ | $n = 10$ (with $21^{\circ}17'$ and $18^{\circ}04'$)[b] | | | | |
| Other researchers according Bošković's source: La Lande: *Astronomie* (de Lalande and Dupius, 1771). | | | | | | |
| Dominique Cassini[c] | $2^{s}08^{\circ}$ | Tom III., book XX. | $7^{\circ}\frac{1}{2}$ | Tom III. | $27\frac{1}{2}$ | Tom III. |
| Dominique Cassini | | | | | $25^{j}10^{h}$ | Tom IV. |
| Cassini (son)[d] | $2^{s}10^{\circ}$ | | | | | |
| L'Isle (1713)[e] | $1^{s}26^{\circ}$ | | $6^{\circ}36'$ | $6^{\circ}35'$ (de Lalande and Dupius, 1771, *§3161*) | | |
| Scheiner (1636) | $2^{s}10^{\circ}$ | | $7^{\circ}$ | $6^{\circ} < i_{Scheiner} < 8^{\circ}$ | | |
| La Lande[f] | $2^{s}18^{\circ}$ | | $7^{\circ}20'$ | Tom V. | | |
| Cagnoli[g] | $2^{s}17^{\circ}50'$ | | $7^{\circ}15'$ | | | |

[a] In Bošković's time, the Zodiac sign $1^{s} = 30^{\circ}$ usually was used.

[b] Bošković's source value $21^{\circ}17'$ is not correct (Boscovich, 1785b, №114), correct value is determined in the present work $21^{\circ}37'$.

[c] Giovanni Domenico Cassini (1625-1712).

[d] Jacques Cassini (1677-1756) is son of Giovani Domenico astronomer and geodesist.

[e] Joseph Nicolas De L'Isle (1688 – 1768) French astronomer, cartographer, geographer and physicist (https://collection.sciencemuseumgroup.org.uk/people/cp100532/joseph-nicolas-de-lisle).

[f] Joseph Jérôme Le François de La Lande (1732-1807).

[g] Andrea Antonio Cagnoli (1743-1816) diplomat of the Republic of Veneto, astronomer and mathematician in Verona. He worked at Brera observatory and he was a professor in Modena. With his friend J. J. La Lande he published in Paris the work *Trigonometria plana e sferica* (1786).

The results of solar rotation elements determination by other researchers, J. D. Cassini (1678), J. Cassini (1740), La Lande (1775) and Delambre (1775) we take as they appear in Wöhl (1978), Table 1. Ruđer Bošković discussed his values with the results of other researchers as presented in parts III (*Tom III.*) and V (*Tom V.*) of *Astronomia* (de Lalande and Dupius, 1771), while the results of Delambre were published later and we reproduce them from the *Encyclopedia Britannica* (Britannica, 1841).

## 5. Discussion

Construction and application of the telescope in astronomy by Galilei in 1609 and invention of logarithms in 1614 opened new astronomical research opportunities, especially observations of the Sun and trigonometric calculations (Britannica, 1841, *History of astronomy*). This led many observers to study the Sun and derive solar rotation elements. We compare Bošković results from this work with historical sources used by Bošković (Table 8), with values from Carrington and Spörer (Table 9), and sources found in the *Encyclopedia Britannica* (Britannica, 1841) (Table 11). The tables were compiled as follows:

1. The solar rotation elements by J. D. Cassini, the older (1678) and J. Cassini, the younger (1740) and J. J. F. La Lande were taken from La Landes's *Astronomia* (de Lalande and Dupius, 1771).
2. The solar rotation elements by Delambre (1775) were taken from the *Encyclopedia Britannica* (Britannica, 1841).
3. The solar rotation elements in Table 8 we took from *§.XIV. Réflexions* (Boscovich, 1785b).
4. We determined the solar differential rotation in the present work using all convenient 1777 observations of Bošković, which were taken from *Opuscule II* (Boscovich, 1785b, *Appendice*).

In the present work we confirmed that Bošković correctly took the values from the sources used in *Opuscule II.* The English source Britannica (1841) did not publish the solar rotation elements determined by Bošković.

### 5.1. The solar rotation elements in La Landes's *Astronomia*

The relative errors computed with $\Omega$ and $i$ of Spörer (1874) and Carrington (1863) are listed in Table 9 supplemented with $\Omega$ and $i$ for the year 1777 of Ruđer Bošković determinations. The Carrington values have been accepted since 1928 (Waldmeier, 1955).

The solar rotation elements in Table 1 in Wöhl (1978) and in Table 8 (present work) were compared with Carrington's $\Omega$ for the analyzed year and Carrington's or Spörer's $i$ using relative errors in percent (%): $R_\Omega = [(\Omega_i - \Omega_{Carr})/\Omega_{Carr}] \cdot 100\%$ and $R_i = [(i_i - i_{Carr/Sp})/i_{Carr/Sp}] \cdot 100\%$ (Table 10).

We can conclude that $\Omega$ and $i$, the ecliptic longitudes of the ascending node $\Omega_6 = 2^s10°21'$ and $\Omega_{136} = 2^s14°03'$, and the solar equator's inclination to the ecliptic $i_5 = 7°44'$ and $i_{136} = 6°49'$, which Bošković determined and were included in the results in Table 1 in Wöhl (1978), are very good:

$$i_5 = 7°44' \approx 7°15' = 7.25° = i_{Carr}, \tag{16}$$

$$i_{136} = 6°49' \approx 6°58'12" = 6.97° = i_{Sp}. \tag{17}$$

**Table 9.** The solar rotation elements: the ecliptic longitude of the ascending node $\Omega$ and the solar equator inclination $i$, angular velocity $\xi$, and sidereal period $T$ determined by Carrington (1863) and Spörer (1874) with the sources in the footnotes (Waldmeier, 1955, Tabelle 12: *Die Rotationelemente der Sonne*). The $\Omega$ values for 1777 were supplemented.

| Solar rotation elements by Spörer and Carrington | | | | | |
|---|---|---|---|---|---|
| Author | $\Omega$ | $i$ | $\xi$ | $T$ | $\Omega_{1777}[°]$ |
| Carrington[a] | $73.6667 + (J - 1850) \cdot 0.01396$ | 7.25° | 14.1844° | $25.380^{d}$ | 72.647620 |
| Spörer[b] | $74.5230 + (J - 1861) \cdot 0.01400$ | 6.97° | 14.2665° | $25.234^{d}$ | 73.347000 |

[a]Carrington, Obs. of the spots of the Sun. London (1863), pp. 221, 244.

[b]Spörer, Beobachtungen von Sonnenflecken. Anklam 1862; Beobahtungen von Sonnenflecken, Publ. XIII der Astron. Gesselschaft, Leipzig 1874; S.-B. Preuß. Akad. Wiss. Berlin 20 (1884).

**Table 10.** Relative errors of $\Omega$ and $i$ (Wöhl, 1978, Table 1) and present work ($\Omega_{Carr}^{1777} = 72.647620°$, $i_{Carr} = 7.25°$, $i_{Sp} = 6.97°$).

| Researcher | Year | $\Omega$ | $\Omega_{Carr}$ | $R_{\Omega}(\%)$ | $i$ | $i_{Carr/Sp}$ | $R_i(\%)$ |
|---|---|---|---|---|---|---|---|
| J. D. Cassini | 1678 | 68° | 71°15'57" | −4.58 | 7.5° | Carrington | 3.33 |
| J. Cassini | 1740 | 70° | 72°07'52.5" | −2.95 | – | – | – |
| La Lande | 1775 | 78.07° | 72°37'11.2" | 7.51 | 7.33° | Carrington | 1.10 |
| Delambre | 1775 | 80.12° | 72°37'11.2" | 10.33 | 7.32° | Carrington | 0.97 |
| Bošković | 1777 | 70°21' | 72°38'51.7" | −3.16 | 7°44' | Carrington | 6.67 |
| Bošković | 1777 | 74°03' | 72°38'51.7" | 1.93 | 6°49' | Spörer | −2.20 |

They have the least relative errors using Carrington's and Spörer's equations: the relative errors of $\Omega$ are $R_{\Omega_6} = -3.16\%$ and $R_{\Omega_{136}} = 1.93\%$ and the corresponding errors of $i$ are $R_{i_5} = 6.67\%$ and $R_{i_{136}} = -2.20\%$ (Table 10). Ruđer Bošković concluded that his result $i_5 = 7°44'$ is more reliable (which Bošković originally called "stronger") than that of others because it is the arithmetic mean of five determinations in *Tab. VI.* (Boscovich, 1785b).

## 5.2. The solar rotation elements in the *Encyclopedia Britannica*

The 1775 results of Delambre (1738–1822) are briefly described in *Sect. V. – Of the Spots of the Sun, his Rotation, and Constitution.* (pp. 778–784) of the Britannica (1841): "... From four different combinations of equations, derived

**Table 11.** The solar rotation elements determined by Delambre (Britannica, 1841).

| | Node. | | | Inclination. | | | Revolution. | | | Synodic Revol. | | |
|---|---|---|---|---|---|---|---|---|---|---|---|---|
| 1 | 80° | 45' | 7" | 7° | 19' | 17" | $25^{d}$ | $0^{h}$ | $17^{m}$ | $26^{d}$ | $4^{h}$ | $17^{m}$ |
| 2 | 79 | 21 | 35 | 7 | 12 | 37 | | | | | | |
| 3 | 80 | 33 | 40 | 7 | 16 | 33 | | | | | | |
| 4 | 79 | 47 | 55 | 7 | 29 | 4 | | | | | | |
| | 80 | 7 | 4 | 7 | 19 | 23 | Diurnal motion 14°.394 | | | | | |

from eleven observations of the same spot, Delambre computed the following table ..." (Table 11).

The ecliptic longitude of the ascending node $\Omega$ = 80°07'04" and the solar equator's inclination $i$ = 7°19'23" are arithmetic means of the four values rounded to the whole angular second: $\overline{\Omega} = \Sigma\Omega_i/n$ = 320°28'14"/4 = 80°07'04.2"= 80.1178333° ≈ 80°07'04"= 80.11777778° ≈ 80.12° and $\bar{i} = \Sigma i_i/n$ = 29°17'31"/4 = 7°19'22.75"= 7.322986111° ≈ 7°19'23"= 7.323055556° ≈ 7.32°. The ecliptic longitude of ascending node $\Omega$ = 80.12° and solar equator's inclination $i$ = 7.32° are like in Table 1 in Wöhl (1978):

$$\overline{\Omega} = \frac{\sum_{i=1}^{4}\Omega_i}{n} = \frac{320°28'17''}{4} = 80°07'04.2'' = 80.11784722° \approx 80°07'04'' = \\ = 80.11777778° \approx 80.12°, \quad (18)$$

$$\bar{i} = \frac{\sum_{i=1}^{4} i_i}{n} = \frac{29°17'31''}{4} = 7°19'22.75'' = 7.322986111° \approx 7°19'23'' = \\ = 7.323055556° \approx 7.32°. \quad (19)$$

The *Encyclopedia Britannica* described the 1630 work on sunspots paths by Christoph Scheiner, using figures (Britannica, 1841, p. 779, Figs 32-38), but they only reviewed Delambre's solar rotation elements. This was published before the works by Carrington (1863) and Spörer (1874). The *Encyclopedia Britannica* mentions Bošković's 1736 short paper on the sunspots (using the reference *Boscovich, De Mac. Sol. Rom. 1736, 4to*), but it does not discuss his *Opuscule II* (Boscovich, 1785b), which contains his completed work on solar spots and in which he presented his successful determination of the solar rotation elements, $\Omega$, $i$, $T'$, and $T''$.

While his ideas about the structure of matter and microscopic forces were popular among physicists at English universities (Whyte, 1961), his results on solar rotation elements seem to be unknown. In this work we presented solar rotation elements of Delambre (1775) and Bošković (1777) in Table 12, and we compared the results in Table 10 with Carrington's $\Omega$ and $i$, where the relative errors are as follows: Bošković has better $\Omega$'s, $|R_{\Omega_B}| < 10.33\%$, but Delambre has better $i$, $R_{i_D} = 0.97\%$.

**Table 12.** The solar rotation elements determined by Delambre (Britannica, 1841) and Boscovich (1785b).

| Researcher (year) | Ω | $i$ | $T'$ | $T''$ |
|---|---|---|---|---|
| Delambre (1775) | $\Omega_4 = 80°07'04''$ | $i_4 = 7°19'23''$ | $25^{\rm d}00^{\rm h}17^{\rm m}$ | $26^{\rm d}04^{\rm h}17^{\rm m}$ |
| | | | Diurnal motion 14°.394 | |
| Bošković (1777) | $\Omega_5 = 70°21'$ | $i_5 = 7°44'$ | $26.77^{\rm d} =$ | $28.89^{\rm d} =$ |
| Bošković (1777) | $\Omega_{136} = 74°03'$ | $i_{136} = 6°49'$ | $26^{\rm d}18^{\rm h}28.8^{\rm m}$ | $28^{\rm d}21^{\rm h}21.6^{\rm m}$ |

## 5.3. The solar rotation elements in Bošković's *§.XIV. Réflexions*

In *§.XIV. Réflexions* (Boscovich, 1785b, №156-№165) Bošković took solar rotation elements from the French source chapter *XX. Livre vingtieme. De la rotation des planetes, & de leurs taches in Astronomia* (de Lalande and Dupius, 1771). In the present work, we checked whether Bošković correctly took the solar rotation elements from La Lande's *Astronomia* (de Lalande and Dupius, 1771). The solar rotation is described in the chapter *De l'Equatteur solaire, & de la rotation du Soleil* (pp. 393–406) with results in paragraphes *§3160*, *§3161*, and *§3162* (de Lalande and Dupius, 1771) (Figure 7) as follows:

1. The solar rotation period by Cassini is $T'' = 27^{\rm j}12^{\rm h}20^{\rm m}$ and $T' = 25^{\rm j}14^{\rm h}8'$, French *j., jour=day*, (*Elémens d'astron.* pag. 105. *Mém. Acad.* 1700 & fuiv) in *§3160.*
2. The solar equator inclination: by Cassini (1678) is $i = 7°\frac{1}{2}$, Scheiner (1626) $i = 7°$ (*Rosa ursina, pag.* 562) and he was convinced that $i$ is never less than 6° nor more than 8°. De L'Isle (1713) $i = 6°35'$ (*Mémoires pour servir, & c. pag.* 178) in *§3161.*
3. The ecliptic longitude of the ascending node by Cassini is $\Omega = 2^{\rm s}8°$ and $\Omega = 2^{\rm s}10°$ from a great number of observations (*Elémens d'astron. pag.* 100), L'Isle from three observations $\Omega = 1^{\rm s}26°$, Scheiner (1626) $\Omega = 2^{\rm s}10°$ the same value like Cassini in *§3162.*

Bošković correctly took all those solar rotation elements from de Lalande and Dupius (1771), except L'Isle's (1713) for which he wrote $i = 6°36'$ (de Lalande and Dupius, 1771, *§3161*) (Figure 7). The solar rotation elements by Scheiner (1636) and L'Isle (1713) were not analyzed in Table 7, because they are not present in the source (Wöhl, 1978, Table 1).

## 5.4. Differential rotation

During their lifetime, sunspots have approximately constant heliographic latitude $b$, and we can use this fact to check for errors. Observations that have large errors and are to far from the average heliographic latitude for that sunspot are most probably erroneous, so we eliminated them (Figure 4). There were no sufficient $(b, \omega)$ pairs for determination of the solar differential rotation law $\omega(b) = A + B \cdot \sin^2 b$ (Equation 10). Figure 5 presents the measured distribution

p. 403 (de Lalande and Dupius, 1771, p. 403):

rotations des planètes.

3160. M. Caſſini avoit trouvé d'abord la révolution moyenne des taches par rapport à la terre 27ʲ $12^h$ 20′, il chercha cette révolution par rapport à un point fixe, en

Eeeij

p. 404 (de Lalande and Dupius, 1771, p. 404):

404 ASTRONOMIE, Liv. XX.

*Révolution des taches, 25 jours 14 heures 8′.*

diſant : 360° + 27° 7′ 8″, mouvement moyen de la terre par rapport aux équinoxes dans l'eſpace de 27ʲ $12^h$ 20′, ſont à 360° comme 27ʲ $12^h$ 20′ ſont à 25ʲ $14^h$ 8′ ; c'eſt la durée de la rotation du ſoleil par rapport aux points équinoxiaux, ( *Elémens d'aſtron. pag.* 105. *Mém. acad.* 1700 & ſuiv ).

*Inclinaiſon de l'équateur, 7°.*

3161. L'ÉQUATEUR ſolaire, ſuivant les anciennes obſervations de M. Caſſini, eſt incliné ſur l'écliptique de 7° $\frac{1}{2}$, comme on le voit dans l'hiſtoire de l'académie, par M. Duhamel, & dans un abrégé d'aſtronomie fait en 1678, & qui eſt encore manuſcrit. Le P. Scheiner ſuppoſoit en 1626 cette inclinaiſon de 7° ( *Roſa urſina, pag.* 562 ), & il aſſure qu'il ne l'avoit jamais trouvée moins de 6°, ni plus de 8 ; M. de l'Iſle en 1713 trouva cette inclinaiſon de 6° 35′, ( *Mémoires pour ſervir, &c. pag.* 178 ) ; mais il n'y employa que trois obſervations, ſeulement pour donner un exemple de ſa méthode ; or pour avoir quelque certitude en pareille matière, il faut néceſſairement pluſieurs comparaiſons d'obſervations priſes 3 à 3, ou combinées toutes enſemble.

*Nœud aſcendant, 2 ſignes 10°.*

3162. LE NŒUD de l'équateur ſolaire ſur l'écliptique étoit à $2^s$ 8° dans le dernier ſiècle ; car M. Caſſini, dans l'abrégé d'aſtronomie que j'ai cité, dit que le pole boréal du ſoleil répond au huitième degré des Poiſſons ; M. Caſſini le fils trouve ce nœud à $2^s$ 10° par un grand nombre d'obſervations, ( *Elém. d'aſtron. pag.* 100 ) ; mais il ne ſeroit, ſuivant les trois obſervations calculées par M. de l'Iſle, qu'à $1^s$ 26° de longitude. Le P. Scheiner en 1626 le trouvoit vers $2^s$ 10°, comme M. Caſſini, puiſqu'il dit qu'au commencement de Décembre la route des taches eſt rectiligne.

*Incertitude à fixer.*

3163. Les trois réſultats que je viens de rapporter quoiqu'un peu différens entre eux, ne ſuffiſent point encore pour nous faire conclure qu'il y ait des changemens ſenſibles dans la poſition de l'équateur du ſoleil ; les obſervations du P. Scheiner n'étoient pas aſſez exactes, celles de M. de l'Iſle n'étoient pas aſſez éloignées de celles de M. Caſſini, pour qu'on puiſſe eſſayer d'en déduire le mou-of angular velocities of the observed sunspots by Bošković in 1777. The sunspots closer to the solar equator rotate faster and the sunspot closer to the pole rotate slower. The gross errors of Bošković's 1777 observations were mapped onto the apparent solar disk (Figure 4). They could be eliminated in further research.

**Figure 7.** The solar rotation elements – the results of J. D. Cassini, J. Cassini, and La Lande in *Astronomia* (de Lalande and Dupius, 1771, *§3160*, *§3161*, and *§3162*).

## 6. Conclusion

This work confirmed the validity of methods of Ruđer Bošković for determination of the solar rotation elements. Bošković's calculations were successfully repeated in the original way and using modernized forms of Bošković's equations.

Before Ruđer Bošković (1777), the solar rotation elements $\Omega$ and $i$ were determined by D. Cassini (the older) (1678) and J. Cassini (the younger) (1740), as well as J. J. F. La Lande (1775) and Delambre (1775) in Table 1 in Wöhl (1978). The Bošković values for $\Omega$ and $i$ are combined with the original Table 1 in Wöhl (1978) and those of other researchers in Table 7. The relative errors of ecliptic longitudes of ascending node from sunspot pairs $R_{\Omega_{6(1900)}} \approx R_{\Omega_{6(1976)}} \approx 3\%$ and from sunspot triples $R_{\Omega_{136(1900)}} \approx R_{\Omega_{136(1976)}} \approx -2\%$ confirm Bošković's results among other results regarding Carrington's values (Table 9).

Delambre's results are the only numerical results in the Britannica (1841). The results of Cassinies (1678 and 1740) and La Lande (1775) are in *Astronomia* (de Lalande and Dupius, 1771), but they are not in the reputable *Britannica*, which was published before the modern results of Spörer (1874) and Carrington (1863)(Waldmeier, 1955).

In the discussion we verified Wöhl's values of solar rotation elements before 1777 in Table 7 (Wöhl, 1978, Table 1): with Delambre (1775) results in Table 11 (Britannica, 1841), and the results of the Cassinies (1678 and 1740) and La Lande (1775) in *Astronomia* (de Lalande and Dupius, 1771) (Figure 7). We also checked the values of solar rotation elements which Ruđer Bošković used in *§.XIV. Réflexions* (Boscovich, 1785b).

We determined the relative errors of $\Omega$ and $i$ regarding Carrington's and Spörer's $\Omega$ and $i$ for the year 1777 in Table 10. The relative errors show the very good quality of original Bošković values: the relative errors of the ecliptic longitudes of the ascending node $\Omega$ are $R_{\Omega_6} = -3.16\%$ and $R_{\Omega_{136}} = 1.93\%$ and the relative errors of the solar equator inclination regarding ecliptic $i$ are $R_{i_5} = 6.67\%$ and $R_{i_{136}} = -2.20\%$.

In 1777 Ruđer Bošković made sufficient observations of four sunspots which were used for solar differential rotation determination. Some observations were eliminated because they had substantial difference/deviation from their mean heliographic latitude $b$, for example, sunspot 2 on 16 September, 1777, in Table 5 and Figure 4. The numerical results of angular velocity rate $\omega_{sid}$ and its heliographic latitude $b$ describe solar differential rotation, where $\omega_{sid}$ decreases with the increasing latitude $b$ (Table 5, Figure 5).

Bošković's 1777 sunspot observations mapped onto the apparent solar disk show gross errors of the expected sunspot trajectories determined a priori before calculation (Figure 4). Further research should eliminate determined gross errors from Bošković's 1777 measurements a posteriori - after calculation. These corrected measurements would be sufficient for a more reliable determination of the solar differential rotation law, parameters $A$ and $B$ in Equation 10. Bošković’s 1777 sunspot position measurements agree within errors with the generally accepted solar differential law, but it is unlikely that he could determine solar differential rotation profile using only his own measurements.

**Acknowledgments** This work has been supported by the *Croatian Science Foundation* under the project *7549 "Millimeter and submillimeter observations of the solar chromosphere with ALMA"* and the project *SOLARNET* funded by the European Commission under Grant Agreement number 824135. We also acknowledge the support from the *Austrian–Croatian Bilateral Scientific Project "Multi-Wavelength Analysis of Solar Rotation Profile"(2022–2023).*

**Author Contribution** M. H. wrote main manuscript text and prepared all figures and tables. All authors reviewed manuscript.